\documentclass[a4paper,12pt]{article}
\usepackage{jcappub}
\usepackage{amsthm,graphicx,multirow}
\usepackage{epsfig}
\usepackage{latexsym, amssymb} 
\usepackage{amsmath}
\usepackage{comment}
\usepackage{hyperref}

{\newcommand{\lsim}{\mbox{\raisebox{-.6ex}{~$\stackrel{<}{\sim}$~}}}
{\newcommand{\gsim}{\mbox{\raisebox{-.6ex}{~$\stackrel{>}{\sim}$~}}}

\begin{document}

\title{Revisiting a pre-inflationary radiation era and its effect on the CMB power spectrum}%

\author[a]{Suratna Das,}%
\author[b]{Gaurav Goswami,}
\author[c]{Jayanti Prasad,}%
\author[b]{Raghavan Rangarajan}

\affiliation[a]{Indian Institute of Technology, Kanpur 208016, India.}
\affiliation[b]{Theoretical Physics Division, Physical Research Laboratory, Navrangpura, Ahmedabad 380009, India.}
\affiliation[c]{Inter-University Centre for Astronomy and Astrophysics, Post Bag 4, Ganeshkhind, Pune 411007, India.}

\emailAdd{suratna@iitk.ac.in, gaugo@prl.res.in, jayanti@iucaa.ernet.in, raghavan@prl.res.in}

\abstract{
We revisit the scenario where inflation is preceded by a radiation era by considering that the 
inflaton too could have been in thermal equilibrium early in the radiation era.  Hence we
take into account not only the effect of a pre-inflationary era on the inflaton mode
functions but also that of a frozen thermal distribution of inflaton quanta. 
We initially discuss in detail the issues relevant to our scenario of a pre-inflationary
radiation dominated era
and then obtain the scalar power spectrum for this scenario.
We find that the  power spectrum is free from infrared divergences.  
We then use the WMAP and Planck data to determine the constraints on the inflaton
comoving `temperature' and on the duration of inflation.  We find that the best fit value of the duration of inflation is
less than 1 e-folding more than what is required to solve cosmological problems, while
only an upper bound on the inflaton temperature can be obtained. }

\maketitle
\section{Introduction}

The inflationary paradigm
\cite{1980PhLB...91...99S,1980ApJ...241L..59K,1981PhRvD..23..347G,
1982PhRvL..48.1220A,1983PhLB..132..317L}
successfully explains not only the spatial flatness, isotropy and
homogeneity of our observed universe, but also the origin of the
density fluctuations in the early universe which give rise to the
large scale structure we observe today.  Inflation, a period of
quasi-exponential expansion of our universe, stretches
(quantum) scalar fluctuations beyond the horizon thereafter freezing their amplitudes.
These fluctuations may then be treated as classical,
and are the source of the gravitational instabilities which later form
the large scale structure of the universe.  Though the existing observational evidence
seems to support the inflationary paradigm, what happened before
inflation is completely unknown.  Could inflation be preceded by a
radiation dominated era?

The possible consequences of a pre-inflationary radiation era have
been studied earlier.  It is well known in the literature
\cite{Powell:2006yg,Marozzi:2011da,Hirai:2002vm,Hirai:2003dh,Hirai:2004kh,
Hirai:2005tn,Hirai:2007ne,Wang:2007ws}
that the
presence of a pre-inflationary radiation era, where one has
`just-enough' inflation, lowers the quadrupole moment of the CMB
temperature anisotropy spectrum. 
\footnote{
Other attempts to explain the low 
power at low CMB multipoles 
involve 
non-trivial 
topologies of the universe 
\cite{doi:10.1142/S0217732387000318,Sokolov:1993mk,Starobinsky:1993yx,
PhysRevLett.71.20,2003MNRAS.343L..95E,2003MNRAS.344L..65U,2004PhRvD..69d3003U,
2006PhRvL..97m1302C,2007PhRvD..75h4034K}, 
bouncing cosmologies \cite{2004PhRvD..69j3520P,2010arXiv1009.3372L},
various inflationary scenarios (e.g. 
hybrid models of inflation \cite{2003PhLB..570..151K},
multi-field inflation \cite{2003PhLB..570..145F},
inflation which takes place in two stages \cite{2009JCAP...01..009J,2010PhRvD..82b3509J,2014arXiv1412.4298C}
just enough inflation
\cite{2003JCAP...09..010C,2008JCAP...01..002N} which could take place in modified gravity 
theories \cite{2004PhLB..583....1K} or preceded by a fast roll phase
\cite{2003JCAP...07..002C}), etc. 
There have also been attempts at providing various other explanations such as non-primordial causes
\cite{2003astro.ph.12124A,2004PhRvL..92i1301M,2004PhRvD..70h3003G,
2010RAA....10..116H,2014JCAP...02..002D} 
and those based on systematic effects \cite{2010arXiv1003.1073L,2013arXiv1307.0001D}
as well as attempts to relate this low power to other anomalies in the CMB
(see e.g. \cite{2009PhRvD..80b3526D}).
}
Lack of power in the CMB quadrupole
is in accordance with observations like COBE \cite{Hinshaw:1996ut},
WMAP \cite{Spergel:2003cb} and PLANCK \cite{Ade:2013uln} despite the
issue of cosmic variance \cite{Cicoli:2014bja}.  
The transition from a
pre-inflationary radiation era to a quasi-exponential inflationary era,
and its effect on the inflaton mode functions,
has been studied (i) when the transition is instantaneous
\cite{Powell:2006yg,Marozzi:2011da,Hirai:2002vm,Hirai:2003dh,Hirai:2004kh,Hirai:2005tn,Hirai:2007ne} 
and (ii) when the transition is continuous \cite{Wang:2007ws}. 
The first approach yields a
`ringing-effect' in the lower multipoles of the $TT$ anisotropy power
spectrum due to the abrupt matching of wave functions at the
transition boundary \cite{Powell:2006yg,Marozzi:2011da,Hirai:2002vm,Hirai:2003dh,Hirai:2004kh,Hirai:2005tn,Hirai:2007ne}, 
while the second method is devoid of any such effect due to a continuous
transition between these two phases \cite{Wang:2007ws}.
The lowering of the CMB quadrupole moment is evident in both these approaches to study
a pre-inflationary radiation era.
Alternatively, in Ref. 
\cite{2006PhRvL..96l1302B}
the authors considered a scenario where the inflaton itself could have
been in thermal equilibrium at some very early epoch possibly near the Planck era.
The effect of this pre-inflationary dynamics
was incorporated by considering a thermal rather than a vacuum state for
the inflaton, i.e., by setting
$\langle a_{\bf k}^\dagger a_{\bf k'}\rangle= [\exp(k/T)-1]^{-1}\delta^3({\bf k}-{\bf k'})$, where $a_{\bf{k}}^\dagger
  a_{\bf{k}}$ is the number operator for the inflaton modes and $T$ is
  the inflaton comoving temperature. In contrast to the studies in Refs. 
 \cite{Powell:2006yg,Marozzi:2011da,Hirai:2002vm,Hirai:2003dh,Hirai:2004kh,Hirai:2005tn,Hirai:2007ne,Wang:2007ws} 
this scenario led to an enhancement in power at low CMB multipoles corresponding to large angular scales.
  This suggests that there exist conflicting effects 
  of a pre-inflationary radiation era: while
  the modified mode functions of the inflaton field
  lower the quadrupole moment, thermal initial conditions on the inflaton quanta tend to
  increase the power for the same.

In this article, we consider both these effects
simultaneously unlike in earlier works that consider only one effect or the other.
{
We find that  
the  effects of the
pre-inflationary era are only effective observationally if inflation
lasts for the bare minimum number of e-folds required to solve the
horizon and flatness problems.  This is similar to the scenario when one
considers a vacuum state with modified mode functions for the inflaton
\cite{Powell:2006yg,Marozzi:2011da,Hirai:2002vm,Hirai:2003dh,Hirai:2004kh,Hirai:2005tn,Hirai:2007ne,Wang:2007ws}  
(indicating that the effect of the thermal state is suppressed by the
effect of the modified mode functions).}
Such `just-enough'
inflationary scenarios can be advocated from the fact that a large
amount of inflation requires some fine-tunning
\cite{Hawking:1987bi,Gibbons:2006pa} and that string landscape models
suffer from the $\eta-$problem \cite{Freivogel:2005vv} which does not
allow them to sustain longer inflation.

The seminal work of Ford and Parker \cite{Ford:1977in}
showed that a pre-inflationary era, radiation or matter, can cure the
infrared divergences which turn up in correlations of inflationary
observables. We regard this as another motivation to study the
consequences of a pre-inflationary radiation era in detail
\cite{Janssen:2009nz,Koivisto:2010pj,Marozzi:2011da}.

We begin in
\textsection \ref{sec:conds} by discussing the assumptions and conditions 
we presume in our analysis.
In \textsection
\ref{sec:evol}, we evaluate the primordial power spectrum of scalar
perturbations 
by matching the inflaton mode functions in the inflationary era with those of the
pre-inflationary radiation era, and by including a thermal distribution for the inflaton.
In
\textsection \ref{sec:param_est}, we use the WMAP and Planck data to
determine the best-fit values or constraints on the  duration of inflation and on the comoving temperature 
of the inflaton thermal distribution.
We then conclude with
a discussion of  various issues 
in
\textsection \ref{sec:conclusions}.

\section{Pre-inflationary radiation era}\label{sec:conds}

In this work we shall assume that before inflation began, the universe
was described by a spatially flat FRW spacetime with small
perturbations and was dominated by a radiation fluid whose equation of
state was of the form $p = \rho/3$.  At some epoch in the early
universe, semiclassical general relativity and quantum field theory
would have become valid and the calculations we shall present are
applicable from this moment onwards.  
Below we discuss 
the
assumptions that are relevant 
to the scenario that we are considering.


\subsection{The little horizon problem}

We presume an FRW metric associated with an isotropic and homogeneous universe
prior to inflation in our analysis of cosmological 
perturbations during the pre-inflationary radiation dominated era.
It requires a level of fine-tuning at the Planck epoch for this assumption
of isotropy and homogeneity to be valid from $t_{\rm Pl}$ to $t_i$ when inflation commences.
The inflationary scenario too requires gradient energy to be sub-dominant on
the horizon scale at the beginning of inflation.

If the Hubble parameter at the beginning of inflation is $H_{i}$ and the scale factor at the beginning of inflation is 
$a_i$, then the physical size $l_1$ of the scale 
corresponding to $H_{i}$ at the Planck time is
\begin{equation}
 l_1 = 
 \frac{ H_{i}^{-1} a (t_{\rm Pl})} {a_i} \; .
\end{equation}
If $l_2$ is the physical size of the horizon at the Planck time, 
then $l_2 = 1/H(t_{\rm Pl})$, and 
\begin{equation}
 \frac{l_1}{l_2} = \frac{a_{\rm Pl} H_{\rm Pl}}{a_i H_i} = \sqrt{ \frac{M_{\rm Pl}}{H_i}} \; .
\end{equation}
So, assuming $H_{i} \sim 10^{-4} M_{\rm Pl}$, 
where $M_{\rm Pl}$ is the reduced Planck mass,
we need to assume that at the Planck epoch, the universe was isotropic and 
homogeneous 
on a length scale which is ${\cal O}(100)$ times larger than the Planck length
for the FRW metric to be valid till $t_i$.
If the energy scale of inflation or $H_i$ is lower, 
one shall require even more fine-tuning at early times.

\subsection{The little flatness problem}
The dimensionless curvature density parameter is defined by $\Omega_K
= -K/(a^2H^2)$, where $K$ could be -1, 0 or +1.  In our analysis below of
perturbations in the pre-inflationary radiation dominated universe we
presume that $\Omega_K$ is negligible.  An upper bound on the curvature is
also required to ensure that the universe does not collapse before the onset
of inflation (if $K=+1$).
Since $\Omega_K$ increases in a decelerating
universe, can we justify ignoring it? 
Let 
$t_{\rm Pl}$ and $t_i$ be the Planck time and the epoch 
when inflation starts. 
Suppose the pre-inflationary radiation era lasts from $t_{\rm Pl}$ to $t_i$, what is the maximum  value of 
$\Omega_K$ at $t_{\rm Pl}$ if we want $\Omega_K$ to be ignorable 
{before} inflation? 

In the pre-inflationary era $\Omega_{r} + \Omega_{\phi} + \Omega_K = 1$,
where $\Omega_{r,\phi}$ refer to the radiation and inflaton field component of the energy density,
and let us assume that we can ignore $\Omega_K$ if it is ${\cal O}(0.01)$, 
i.e., at the onset of inflation, we expect that $\Omega_{r} \approx \Omega_{\phi} \approx {\cal O}(1)\gg\Omega_K$.
Then, since $\Omega_K \sim a^{-2}$ 
for a radiation dominated universe
\begin{equation}
 \Omega_K(t_{\rm Pl}) \approx \Omega_K (t_i) \left(\frac{H_i}{M_{\rm Pl}} \right) \; ,
\end{equation}
which, assuming $H_{i} \sim 10^{-4} M_{\rm Pl}$, 
turns out to be $10^{-6}$.
If the energy scale of inflation is lower, the fine-tuning problem  
gets more serious.

\subsection{Local thermodynamic equilibrium}
\label{sec:LTE}

The stress tensor can be evaluated for any collection of particles but when the distribution function 
of the collection of particles is close to its form in thermodynamic equilibrium, the fluid approximation is valid
and the stress tensor takes the simple form we use in the Friedmann equations in cosmology.
In a pre-inflationary era, is the fluid approximation valid? 
Was there enough time before inflation so that enough collisions could have happened that caused local 
thermodynamic equilibrium?
That is, in the pre-inflationary era, how does the mean free path compare with the Hubble distance?

If a marginal coupling $g$ with a vertex with three external lines contributes to 2-2 scattering process 
which sets the equilibrium, then, assuming that the coupling $g \approx {\cal O}(10^{-1/2})
$
(the typical value of the Standard Model gauge couplings at the Planck scale), the ratio 
\begin{equation}
 \frac{t_{\rm coll}}{t_{\rm Hubble} } \sim \frac{T}{g^4 M_{\rm Pl}} \; ,
\end{equation}
is smaller than unity only when the ``temperature'' is hundred times smaller than the Planck mass
which corresponds to $H/M_{\rm Pl} \sim 10^{-4}$.
If the Hubble parameter before inflation 
is much smaller than this, pre-inflationary relativistic particles 
do get enough time to attain thermal equilibrium and
the fluid limit; if not, the fluid approximation is not valid
\footnote{However 
if we lower the energy scale of inflation the fine-tuning required to
have a pre-inflationary radiation dominated universe described by spatially flat FRW universe shall increase, 
unless the universe has a non-trivial topology \cite{2004JCAP...10..004L}.
(Quantum creation of a universe with non-trivial topology has been considered
in Ref. \cite{Zeldovich:1984vk}.)}.

As an important aside, let us see what happens if the equilibrium is set by a gravitational interaction.
The leading gravitational interaction is a dimension five operator 
{coupling
a graviton to, say, two scalars}
and is suppressed by 
the Planck mass
\begin{equation}
 {\cal L}_{\rm int} \supseteq c \frac{{\cal O}_5}{M_{\rm Pl}} \; ,
\end{equation}
which implies $\sigma(E) \sim c^4 E^2/M_{\rm Pl}^4$ for 2-2 scattering mediated by a graviton.
Hence 
\begin{equation}
 \frac{t_{\rm coll}}{t_{\rm Hubble} } \sim \frac{M_{\rm Pl}^3}{T^3 c^4} \; ,
\end{equation}
so that the mean collision time is smaller than the Hubble time only for super-Planckian temperatures.
Thus, gravity mediated interactions can cause thermal equilibrium at sub-Planckian temperatures only 
if $c > 1$.

Further exploring the scenario where $t_{\rm coll}>t_{\rm Hubble}$ in the pre-inflationary radiation era,
we could assume that somehow at the Planck time the universe was in local 
thermodynamic equilibrium but soon went out of equilibrium. Then, the distribution function of the relativistic particles is frozen in the equilibrium form while the relativistic particles form a hot decoupled relic radiation.  
Can the perturbations around the equilibrium distribution function be treated in 
the fluid approximation?
Had the particles forming the pre-inflationary stuff been non-relativistic (like CDM), the fluid approximation 
would still have been valid for perturbations and we could have described them by just two variables: the density 
contrast $\delta$ and the peculiar velocity ${\bf v}$ (see, e.g., Sec. 4.5 of Ref. \cite{2003moco.book.....D}). 
But they are relativistic, and so the fluid approximation does not hold good for the perturbations. 
In particular, the ideal fluid approximation breaks down for them as they cause anisotropic stresses 
(see Eq. 5.33 of Ref. \cite{2003moco.book.....D}). Note that for photons and massless neutrinos at decoupling, 
the anisotropic stress is small because at the epoch of decoupling, the universe was already 
matter 
dominated.

Below we assume that the scale of inflation is low enough that the radiation has sufficient time to thermalise before inflation
commences and so the fluid approximation is valid.

\subsection{Small $\Phi$}
 \label{sec:other_assumptions}

To evaluate the scalar primordial power spectrum, we need to find the evolution of the comoving curvature perturbation $\cal R$, 
which during inflation takes 
the form
\begin{equation} \label{eq:curv_pert_infl}
 {\cal R} = \Phi + \frac{H}{\dot{\phi_0}} \delta \phi \,,
\end{equation}
where $\Phi$ is the metric perturbation and 
the inflaton field $\phi(\mathbf{x},t)$ is
decomposed into a ``classical'' part $\phi_0(t)$ (which is its
background value) and a quantum fluctuating part
$\delta\phi(\mathbf{x},t)$ as 
\begin{eqnarray}
\phi(\mathbf{x},t)=\phi_0(t)+\delta\phi(\mathbf{x},t).
\label{phidecomp}
\end{eqnarray}
For now we work in the conformal Newtonian gauge.
During inflation, the metric perturbation $\Phi$ is negligible as compared to $\delta \phi$
(see Fig. (6.8) and Sec. 6.5.2 of Ref. \cite{2003moco.book.....D}) and it becomes non-negligible only as inflation ends.
Furthermore,
for 
a radiation dominated universe,
\cite{mukhanov2005physical}
 \[
    \Phi_k \sim 
\begin{cases}
    {\rm constant} ,& {\rm for}~~  ({\rm super-Hubble}), \\
    \frac{\sin x}{x^2}, & {\rm for}~~  ({\rm sub-Hubble}) \; ,
\end{cases}
\]
where $x = k \tau$ and $\tau$ is the conformal time.
So we presume that $\Phi_k$ of the pre-inflationary radiation era dies down.  
This may not be strictly true for modes
that enter the horizon just before inflation begins -- for such modes we presume the super-Hubble value of $\Phi_k$
is small.  Then the contribution of any pre-inflationary $\Phi_k$ can be ignored during inflation.
Given the above, the scalar power spectrum during inflation
is determined by the quantum fluctuations of $\delta \phi$ 
only and this is the quantity whose evolution we follow in the next section.
In \textsection \ref{sec:evol}, we shall work with a gauge invariant variable $\delta\varphi^{\rm gi}$ 
which equals the field fluctuation in the conformal Newtonian gauge and shall perform matching of this variable at 
the transition from a radiation dominated universe to an inflaton dominated universe.  
We will find the power spectrum
of $\cal R$ from the power spectrum of $\delta \phi^{\rm gi}$.
We shall also assume an instantaneous transition from a pre-inflationary radiation
era to an inflationary era.

\subsection{Initial conditions for the mode functions}\label{eq:expected_ps}

{When we have a pre-inflationary radiation dominated era, at early enough times, 
modes that are outside the horizon during the radiation era 
become 
subhorizon as time passes.  The modes of cosmological interest enter
the horizon before the inflationary era commences.
We apply initial conditions corresponding to plane waves with a positive frequency 
for these modes, and also argue below that this is  
justified for modes that enter
the horizon at the very end of the radiation dominated era.
} 

\subsection{Thermal initial state}  
\label{subseq:Thermalinitcond}

If there is a pre-inflationary radiation dominated era, apart from the change in the inflaton
mode functions,
the state of the inflaton quanta could also be modified due to thermal effects as we now argue. 

We may picture the energy density of the pre-inflationary universe as including contributions from 
(i) a species of relativistic particles (which form a fluid with an equation of state of the form $p = \rho/3$ with 
$\rho$ falling as $a^{-4}$), and (ii) a coherent scalar field $\phi$ whose energy density does not dilute.
Then, at some stage, the energy density of the radiation falls below the energy density of the scalar field and 
the universe begins inflating. 
One can assume \cite{2006PhRvL..96l1302B} that the quanta of the inflaton fluctuations
$\delta \phi$ 
decouples from the rest of the plasma at some time $t_{\rm d}$ before inflation begins. 
After decoupling, the quanta of $\delta \phi$ travel along geodesics in the spacetime 
so that the distribution function $f(t,{\bf x},{\bf p})$ is conserved (just like for collisionless dark matter;
also, notice that $\bf p$ is the physical momentum).
Assuming that the decoupling happens suddenly
at a temperature ${\cal T}_{d}$, the frozen distribution function 
is the equilibrium distribution function $f_{\rm eq}$ 
at
the 
epoch of decoupling: 
$f_{\rm d} = f_{\rm eq} (t_{\rm d},p_{\rm d})$, where, $p_{\rm d}$ is the 
physical momentum of the particle at the epoch of decoupling.
Then, for the essentially non-interacting gas of (nearly) massless inflatons, the 
distribution function after decoupling 
is given by
\begin{equation}
f(t,{\bf x},{\bf p}) = \frac{1}{ \exp \left( {\frac{a(t) p(t)}{ a(t_{\rm d}) {\cal T}(t_{\rm d}) } } \right) - 1} \; ,
\end{equation}
which has the same form as the equilibrium mean occupation number for a relativistic species with the temperature
\begin{equation}
 {\cal T}(t) = \frac{{\cal T}(t_{\rm d}) a(t_{\rm d})}{a(t)} \; ,
\end{equation}
even though the species $\delta \phi$ has fallen out of equilibrium. Notice that, just like for 
any decoupled species, this \textquotedblleft temperature\textquotedblright~ falls strictly as $a^{-1}$ 
unlike the temperature of a species which is in equilibrium (for which the relation between $\cal T$ and $a$
depends on the number of relativistic degrees of freedom). Defining the comoving temperature $T$ by 
$T = a(t) {\cal T}(t) = {\cal T}(t_{\rm d}) a(t_{\rm d})$, the comoving temperature can be constrained
\cite{2006PhRvL..96l1302B}, as explained below.

In this scenario, the modes of the quantum field $\delta \phi$ are not expected to be in a vacuum state but in 
a thermal state.  This causes the scalar Primordial Power Spectrum 
to become \cite{2006PhRvL..96l1302B}
\begin{equation}
  P_{\cal R} (k) = A_s \left( \frac{k}{k_P} \right)^{n_s - 1} 
    \coth \left( \frac{k}{2  T}\right) \; ,  
\end{equation}
and $T$ being the comoving temperature introduces a length scale in 
the power spectrum.
The observable $k$ range is taken to be $10^0 ~{\rm Mpc}^{-1}$ to $10^{-5} ~{\rm Mpc}^{-1}$ 
and 
the presence of the coth factor in the above equation increases the power in CMB anisotropies at large angular scales.
In order to not substantially affect the power spectrum over the observable $k$ range, one requires the denominator 
in the coth function to be smaller than the present Hubble scale $H_0$, this constrains the comoving temperature 
\cite{2006PhRvL..96l1302B}
 \begin{equation}
  T \le 4.2 H_0 \approx 10^{-3} {\rm Mpc}^{-1} \; .
 \end{equation}
 Let the mode $k_0$ be such that $k_0 = a_0 H_0$ (i.e. it is crossing the Hubble radius today). 
 If this mode exited the Hubble radius during inflation at an epoch $t_*$, then 
 $T = {\cal T}_{*} a_{*}$, where, ${\cal T}_{*}$ and $a_{*}$ denote the physical temperature and scale factor 
 at $t=t_*$ when the Hubble parameter is $H_*$. Then, the above relation implies that ${\cal T}_{*} \le 4.2 H_{*}$, 
 using $a_{*} H_{*}= a_0 H_0$ (and setting $a_0 = 1$).
 At the onset of inflation $\rho_r = \rho_\phi$ where $\rho_\phi \approx V$ (slow-roll approximation)
 and $\rho_r = \frac{\pi^2 g_*}{30} {\cal T}_{\gamma}^4$ ($g_*$ is the number of relativistic species in the 
 pre-inflationary plasma and ${\cal T}_\gamma$ is the physical temperature of the pre-inflationary radiation 
 at the epoch of onset of inflation). Assuming the physical temperature of the inflaton at the 
 onset of inflation ${\cal T}_i = {\cal T}_\gamma$ 
 \footnote{This assumption is valid when there is no entropy production between $t_{d}$ and $t_i$.}, we get
 \begin{equation} \label{eq:Ti}
  {\cal T}_i = \left( \frac{30}{\pi^2 g_*} \right)^{1/4} V^{1/4} \,.
 \end{equation}
 The condition ${\cal T}_{*} \le 4.2 H_{*}$ implies, using the fact that
 $H_{*} = \left( \frac{V^{1/4}}{M_{\rm Pl}} \right) V^{1/4}$,
  \begin{equation} \label{eq:T0}
 {\cal T}_{*} \ll V^{1/4} \;.
 \end{equation}
 Eqs. (\ref{eq:Ti}) and (\ref{eq:T0}) together imply that 
 in order to not affect the CMB temperature anisotropies at large angular scales,  one needs more than the minimum 
 amount of inflation (determined by when the mode $k_0$ left the Hubble radius during 
 inflation) 
 in such a scenario
 \cite{2006PhRvL..96l1302B}.
 (Moreover, the amount of power at low $\ell$ values in 
 the temperature anisotropy and
 B-mode polarization of CMB 
 also increases if gravitons too were in thermal equilibrium at
 the Planck era
 \cite{1993PhRvD..48..439G,2006PhRvL..97y1301B,Zhao:2009pt}.)

 While
 the constraint according to Ref. \cite{2006PhRvL..96l1302B} 
 on the comoving temperature of the inflaton quanta is $T \lesssim  10^{-3} {\rm Mpc}^{-1}$, 
 the detailed analysis of \textsection \ref{sec:param_est} in the present work has improved this constraint to 
 $T \lesssim  10^{-4} {\rm Mpc}^{-1}$.


\section{Evolution of perturbations} \label{sec:evol}

Having set up the basic scenario in \textsection \ref{sec:conds}, one now has to solve for the
metric perturbations 
which we present here in a gauge invariant form. 
We follow Ref.~\cite{1992PhR...215..203M} 
for our analysis. 
Considering only the scalar perturbations of the metric, the most general form of the perturbed 
spatially flat metric in conformal coordinates takes the form
\begin{eqnarray}
g_{\mu\nu}\equiv g^0_{\mu\nu}+\delta g_{\mu\nu}=a^2(\tau)\left(
\begin{array}{ccc}
1+2A & & -\partial_iB \\
-\partial_iB & & -\left(1-2\psi\right)\delta_{ij}-2\partial_i\partial_jE
\end{array}
\right),
\label{per-met}
\end{eqnarray}
%
As these scalar perturbations
are not gauge invariant quantities, it is useful to construct
gauge invariant variables (known as Bardeen potentials) out of these
metric perturbations as
\begin{eqnarray}
\Phi&=&A+{\mathcal H}(B-E^\prime)+(B-E^\prime)^\prime, \label{gi-Phi} \\
\Psi&=&\psi-{\mathcal H}(B-E^\prime), \label{gi-Psi}
\end{eqnarray}
where ${\mathcal H}=\frac{a^\prime}{a}=aH$ and
$X^\prime\equiv\frac{\partial X}{\partial\tau}$.  Absence of
anisotropic stress in the stress-energy tensor puts a constraint on
these gauge invariant quantities yielding $\Phi=\Psi$, irrespective of
any particular choice of gauge.  This is the quantity $\Phi$ introduced
in \textsection\ref{sec:other_assumptions}.

The fluctuation $\delta\phi(\mathbf{x},t)$ in Eq. (\ref{phidecomp}) is not a
gauge invariant quantity and a gauge invariant perturbation of the
inflaton field $\delta\varphi^{\rm gi}(\mathbf{x},t)$ can be constructed with
the metric fluctuations as
\begin{eqnarray}
\delta\varphi^{\rm gi}=\delta\phi+\phi_0^\prime(B-E^\prime).
\label{gi-phi}
\end{eqnarray}
In this perturbed background the equation of motion of the
gauge invariant inflaton perturbation, taking into account
$\Phi=\Psi$, is
\begin{eqnarray}
\delta\varphi^{\rm gi\prime\prime}+2aH\delta\varphi^{\rm gi\prime}-\nabla^2\delta\varphi^{\rm gi}+V,_{\phi\phi}a^2\delta\varphi^{\rm gi}=4\phi_0^\prime\Phi^\prime-2V,_\phi a^2\Phi\,.
\label{phi-gi-eom}
\end{eqnarray}
In momentum space, we have
\begin{eqnarray}
\delta\varphi_k^{\rm gi\prime\prime}+2\mathcal{H}\delta\varphi_k^{\rm gi\prime}+k^2\delta\varphi_k^{\rm gi}+V,_{\phi\phi}a^2\delta\varphi_k^{\rm gi}=4\phi_0^\prime\Phi_k^\prime-2V,_\phi a^2\Phi_k.
\label{phik-gi-eom}
\end{eqnarray}
Thermal effects can modify the equation of motion for the homogeneous and non-zero momentum modes by generating additional terms in the effective potential \cite{Vilenkin198291,Vilenkin:1982wt,Starobinsky1982175,1986LNP...246..107S}, even with the frozen thermal distribution of inflaton quanta.  One expects a mass correction, for example, 
$\sim\lambda T^2/a^2$ for a $\lambda\phi^4$ potential.  (Recall that $T/a$ is the physicaltemperature.)
We compare this with $k^2/a^2$.  For modes within the horizon at the beginning of inflation $k\ge a_i H_i$ and 
$k/a_i \ge H_i \sim {\cal T}_i^2/M_{\rm Pl}>{\sqrt{\lambda}} {\cal T}_i$ if $\lambda$ is sufficiently small as we shall presume.  Then for these modes of interest one can ignore these thermal correction to the potential in the equations of motion.  Nevertheless one can treat the potential $V$ above as including
thermal corrections.


\subsection{Subhorizon primordial perturbations in the pre-inflationary radiation era}

The evolution of the scalar field perturbations in the pre-inflationary radiation era is governed by Eq.~(\ref{phik-gi-eom}).
From the discussion in \textsection \ref{sec:other_assumptions} we ignore the
RHS of 
Eq.~(\ref{phik-gi-eom}).
For a radiation dominated universe, $a=a_1 \left( \frac{t}{t_1} \right)^{1/2}$ and
so the scale factor in terms of conformal time is 
\begin{equation}
 a = \left[ \tau + \tau_1 - \frac{2t_1}{a_1} \right] \frac{a_1^2}{2t_1} \; .
\end{equation}
If we now set $\tau_1$ to $\frac{2t_1}{a_1}$ and choose $t_1$ to represent the epoch of transition from 
the pre-inflationary radiation era to
inflation, i.e. $t_i$, then
\begin{equation}
 a = \frac{a_i^2}{2t_i} \tau \; .
 \label{atau}
\end{equation}
We restrict ourselves to a model of inflation in which the second 
potential slow-roll parameter $\eta_V$ (defined to be $M_{\rm Pl}^2 V_{\phi \phi}/V$) is negative as, for
example, for a 
potential of the form $V(\phi)=V_0-m^2\phi^2/2$.  (This allows us to recast the equations below in
a more convenient form for solving.  This assumption does not affect our results as mentioned 
in \textsection \ref{justification}.)
One can then write Eq.~(\ref{phik-gi-eom})  as
\begin{eqnarray} \label{eq:delta_phi_rad_era}
\delta\varphi_k^{\rm gi\prime\prime}+2{\cal {H}}
\delta\varphi_k^{\rm gi\prime}+(k^2-\tilde c^2\tau^2)\delta\varphi_k^{\rm gi}=0,
\label{deltaphigieom}
\end{eqnarray}
where 
\begin{equation}
\tilde c^2=-\eta_V \frac{V}{M_{\rm Pl}^2} \frac{a_i^4}{4t_i^2}\,.
\end{equation}
Below we shall take
$\tilde c$ to be positive. Redefining the field as
$\chi_k=a\delta\varphi_k^{\rm gi}$ one gets
\begin{eqnarray}
\chi_k^{\prime\prime}+(k^2-\tilde c^2\tau^2)\chi_k=0.
\end{eqnarray}
We rewrite the above equation as 
\begin{eqnarray}
\frac{d^2\chi_k}{dz^2}+\left(\nu+\frac12-\frac14z^2\right)\chi_k=0,
\end{eqnarray}
where $\nu+\frac12=-\frac{k^2}{2\tilde c}$ and $z=i\sqrt{2\tilde c}\,\tau$ 
and considering $\tilde a=-(\nu+\frac12)=\frac{k^2}{2\tilde c}$ we get 
\begin{eqnarray}
\frac{d^2\chi_k}{dz^2}-\left(\frac14z^2+\tilde a\right)\chi_k=0 \; ,
\end{eqnarray}
(compare with, e.g., Eq. (19.1.2) of Ref. \cite{abramowitz1964handbook}).
The even and odd solutions of the above equation are given in Eqs. (19.2.5)
and (19.2.6) of Ref. \cite{abramowitz1964handbook} which are as follows
\begin{eqnarray}
\chi_k^1&=&1+\tilde{a}\frac{z^2}{2!}+\left(\tilde{a}^2+\frac12\right)\frac{z^4}{4!}+\left(\tilde{a}^3+\frac72\tilde{a}\right)\frac{z^6}{6!}+\left(\tilde{a}^4+11\tilde{a}^2+\frac{15}{4}\right)\frac{z^8}{8!}\nonumber\\
&&+\left({\tilde{a}^5+25\tilde{a}^3+\frac{211}{4}\tilde{a}}\right)\frac{z^{10}}{10!}+\cdots\nonumber\\
\chi_k^2&=&z+\tilde{a}\frac{z^3}{3!}+\left(\tilde{a}^2+\frac32\right)\frac{z^5}{5!}+\left(\tilde{a}^3+\frac{13}{2}\tilde{a}\right)\frac{z^7}{7!}+\left(\tilde{a}^4+17\tilde{a}^2+\frac{63}{4}\right)\frac{z^9}{9!}\nonumber\\
&&+\left({\tilde{a}^5+35\tilde{a}^3+\frac{531}{4}\tilde{a}}\right)\frac{z^{11}}{11!}+\cdots
\end{eqnarray}
Hence the asymptotic forms of the above two solutions with $\tilde a$ large are
\begin{eqnarray}
\chi_k^1&\approx&1+\tilde{a}\frac{z^2}{2!}+\tilde{a}^2\frac{z^4}{4!}+\tilde{a}^3\frac{z^6}{6!}+\tilde{a}^4\frac{z^8}{8!}+\tilde{a}^5\frac{z^{10}}{10!}+\cdots
=\cosh(\sqrt{\tilde a}z)\nonumber\\ 
\chi_k^2&\approx&z+\tilde{a}\frac{z^3}{3!}+\tilde{a}^2\frac{z^5}{5!}+\tilde{a}^3\frac{z^7}{7!}+\tilde{a}^4\frac{z^9}{9!}
+\tilde{a}^5\frac{z^{11}}{11!}+\cdots=\frac{1}{\sqrt{\tilde{a}}}\sinh(\sqrt{\tilde a}z)
\label{asympchik}
\end{eqnarray}
If we now define $v_1$ and $v_2$ by
\begin{eqnarray}
v_1&=&\chi_k^1+\sqrt{\tilde{a}}\chi_k^2\approx e^{\sqrt{\tilde a}z}=e^{ik\tau}\nonumber\\
v_2&=&\chi_k^1-\sqrt{\tilde{a}}\chi_k^2\approx e^{-\sqrt{\tilde a}z}=e^{-ik\tau},
\end{eqnarray}
the expression for $\chi(k)$ during the radiation era is
\begin{eqnarray}
\chi_k(\tau)=c_1(k)v_1+c_2(k)v_2,
\end{eqnarray}
which in the sub-horizon limit becomes
\begin{eqnarray}
\chi_k(\tau)\approx c_1(k)e^{ik\tau}+c_2(k)e^{-ik\tau} \; .
\end{eqnarray}
To obtain the Minkowski spacetime solutions in the sub-horizon limit, we
choose $c_2(k)=\frac{1}{\sqrt{2k}}$ and $c_1(k)=0$ 
(for all $k$).
Then, for sub-horizon modes
\begin{eqnarray}
\delta\varphi^{\rm gi}_{\rm rad}(k,\tau)
\approx
\frac{1}{a(\tau)\sqrt{2k}}e^{-ik\tau}
\label{phi-rad}
\end{eqnarray}

\subsubsection{Justification for large $\tilde{a}$}\label{justification}

To obtain the asymptotic expressions in Eq. (\ref{asympchik}) we assumed
that $\tilde a$ is large. 
Since $\tilde{a} = \frac{k^2}{2\tilde c}$, 
this is equivalent to assuming that $k^4 \gg 4 {\tilde c}^2$, or
\begin{eqnarray}
 k^4 \gg 4 |\eta_V| \frac{V}{M_{\rm Pl}^4} \frac{a_i^4}{4t_i^2} M_{\rm Pl}^2 \; .
\end{eqnarray}
Since $t = 1/(2H)$ for a radiation dominated universe, the above gives 
\begin{eqnarray}
 k^4 \gg 4 |\eta_V| \frac{V}{M_{\rm Pl}^4} a_i^4 H_i^2 M_{\rm Pl}^2 \; .
\end{eqnarray} 
We need to consider modes for which $k \gsim a_i H_i$, i.e. modes that enter the horizon during the 
pre-inflationary radiation era.  For the smallest $k=a_i H_i$ the above condition then gives 
\begin{eqnarray}
 H_i^2 \gg 4 |\eta_V| \frac{V}{M_{\rm Pl}^2} \; .
\end{eqnarray}
At the epoch of transition 
 $H_i^2 = \frac{V}{3 M_{\rm Pl}^2} $,
which implies that the condition for having large $\tilde a$ is that 
\begin{equation} \label{eq:eta_V_condition}
| \eta_V| \ll \frac{1}{12} \; .
\end{equation}
Thus, if the potential is sufficiently flat, considering $\tilde a$ to be large can be justified
for $k\gsim a_i H_i$.  This analysis justifies the plane wave form of the mode functions at $t_i$ even for
modes which were just entering the horizon at the onset of inflation.

Arguments similar to the above can be used to show that for a potential with $\eta_V$ of either sign,
$k^2\gg \tilde c \tau^2$ in Eq. (\ref{deltaphigieom}) 
at $\tau_i$
for modes of interest  if $| \eta_V| \ll \frac{1}{3}$, thereby
justifying the plane
wave form of the mode functions in the pre-inflationary era at $t_i$.

\subsection{Evolution of primordial perturbations during inflation}
\label{primpertinfl}

Using the background equation for the slow-rolling inflaton field, i.e.
$V,_\phi a^2\approx-2\mathcal{H}\phi_0^\prime$, Eq. (\ref{phik-gi-eom}) can
be written as
\begin{eqnarray}
\delta\varphi_k^{\rm gi\prime\prime}+2\mathcal{H}\delta\varphi_k^{\rm gi\prime}+k^2\delta\varphi_k^{\rm gi}+V,_{\phi\phi}a^2\delta\varphi_k^{\rm gi}=4\phi_0^\prime\Phi_k^\prime+4\mathcal{H}\phi_0^\prime\Phi_k\,.
\label{phik-gi-eom-1}
\end{eqnarray}
The conformal time in quasi-de Sitter space is
$\tau=-\frac{1}{(1-\epsilon)\mathcal{H}}$,
where
the Hubble slow-roll parameter
$\epsilon\equiv-\frac{\dot H}{H^2}=4\pi G\frac{\phi_0^{\prime
    2}}{\mathcal{H}^2}$. 
Now the $0-i^{\rm th}$ component of the Einstein equation 
can be written during inflation as 
\begin{eqnarray}
\Phi^\prime+\mathcal{H}\Phi=4\pi G \phi_0^\prime\delta\varphi^{\rm gi}.
\end{eqnarray}
Using this in the R.H.S. of Eq.~(\ref{phik-gi-eom-1}) yields 
\begin{eqnarray}
\delta\varphi_k^{\rm gi\prime\prime}+2\mathcal{H}\delta\varphi_k^{\rm gi\prime}+k^2\delta\varphi_k^{\rm gi}+V,_{\phi\phi}a^2\delta\varphi_k^{\rm gi}=16\pi G \phi_0^{\prime 2}\delta\varphi_k^{\rm gi}.
\label{phik-gi-eom-2}
\end{eqnarray}
Now, we can relate the last two terms in the above equation to the
slow-roll parameters. 
The Hubble slow-roll parameter
$\eta\equiv-\frac{\ddot\phi_0}{H\dot\phi_0}$
and
the standard slow-roll parameters are $\epsilon_V\equiv\frac{M_{\rm
    Pl}^2}{2}\left(\frac{V'}{V}\right)^2$ and $\eta_V= M_{\rm
  Pl}^2\left(\frac{V''}{V}\right)=\frac13\frac{a^2V,_{\phi\phi}}{\mathcal{H}^2}$. In
the slow-roll regime we have $\epsilon_V\approx\epsilon$ and
$\eta_V\approx\eta+\epsilon$. Hence we can write the above
equation as
\begin{eqnarray}
\delta\varphi_k^{\rm gi\prime\prime}+2\mathcal{H}\delta\varphi_k^{\rm gi\prime}+k^2\delta\varphi_k^{\rm gi}+(3\eta-\epsilon)\mathcal{H}^2\delta\varphi_k^{\rm gi}=0.
\label{phik-gi-eom-3}
\end{eqnarray}
It is convenient to redefine the field as
$\chi_k=a\delta\varphi_k^{\rm gi}$, as we did before, whose Wronskian
yields $\chi_k\chi_k^{\ast\prime}-\chi_k^{\ast}\chi_k^\prime=i$ or
$\chi_k\dot\chi_k^{\ast}-\chi_k^{\ast}\dot\chi_k=\frac{i}{a(t)}$. The
equation of motion for $\chi_k$ during inflation is
\begin{eqnarray}
\chi_k^{\prime\prime}+\left[k^2-\frac{a^{\prime\prime}}{a}+(3\eta-\epsilon)\mathcal{H}^2\right]\chi_k=0.
\end{eqnarray}
In a quasi-de Sitter space
$\frac{a^{\prime\prime}}{a}\approx\frac{1}{\tau^2}(2+3\epsilon)$ and
$\mathcal{H}=-\frac{1}{\tau(1-\epsilon)}$. Thus in a quasi-de Sitter
space the above equation can be written as
\begin{eqnarray}
\chi_k^{\prime\prime}+\left[k^2-\frac{1}{\tau^2}(2-3\eta+4\epsilon)\right]\chi_k=0,
\end{eqnarray}
or as 
\begin{eqnarray}
\chi_k^{\prime\prime}+\left[k^2-\frac{1}{\tau^2}\left(\nu_\chi^2-\frac14\right)\right]\chi_k=0,
\end{eqnarray}
where $\nu_\chi^2\equiv\frac94-3\eta+4\epsilon$. The solution of the
above equation can be written as
\begin{eqnarray}
\chi_k=\sqrt{-\tau}\left[\tilde c_1(k)H_{\nu_\chi}^{(1)}(-k\tau)+\tilde c_2(k)H_{\nu_\chi}^{(2)}(-k\tau)\right],
\end{eqnarray}
where $H_{\nu_\chi}^{(1)}$ and $H_{\nu_\chi}^{(2)}$ are the Hankel
functions of the first and second kind.

Thus during the inflationary era the mode functions of the gauge invariant
inflaton fluctuations have a solution
\begin{eqnarray}
\delta\varphi^{\rm gi}_{\rm inf}(k,\tau)\equiv a(\tau)^{-1}\chi_k=a(\tau)^{-1}\sqrt{-\tau}\left[\tilde c_1(k)H_{\nu_\chi}^{(1)}(-k\tau)+\tilde c_2(k)H_{\nu_\chi}^{(2)}(-k\tau)\right].
\label{phi-inf}
\end{eqnarray}
(Had we ignored the R.H.S. of Eq. (\ref{phik-gi-eom-1}) due to small $\Phi_k$, we would have obtained
$\nu_\chi^2\equiv\frac94-3\eta+3\epsilon$.)

\subsection{Mode function matching and the power spectrum} \label{sec:matching}
For a transition from a radiation dominated era to an inflationary era
at $t=t_i$ the form of the scale factor changes as
\begin{eqnarray}
a(t)&=&a_i(t/t_i)^{1/2}, \quad\quad t\leq t_i \\
a(t)&=&a_ie^{H_i(t-t_i)+\frac{\dot H_i}{2}(t-t_i)^2}, \quad\quad t>t_i.
\end{eqnarray}
$H_i$ is the scale factor at the time of the transition.
$H_i=\frac{1}{2t_i}$. $a$ and $\dot a$ are continuous at $t_i$.
We have seen in the previous two sub-sections that the evolution of the
gauge invariant inflaton fluctuations during the pre-inflationary radiation
era and in the inflationary era is given by Eq.~(\ref{phi-rad}) and
Eq.~(\ref{phi-inf}). 
We define
$z_{R,I}\equiv
-k\tau=\frac{k}{(1-\epsilon_{R,I})aH}$ 
where $\epsilon_{R,I}$ are the slow-roll parameter during the radiation and inflationary eras.
$\epsilon_R=2$ while $\epsilon_I\ll1$. 
$\tau$ in the definition of $z_R$ above is consistent with Eq. (\ref{atau}).
$z(k)$ is not continuous at
the transition (unlike in some earlier works such as Refs. \cite{
Hirai:2002vm, Hirai:2003dh, Hirai:2004kh, Hirai:2005tn, Hirai:2007ne}
thereby
 giving somewhat different final expressions).
Then
\begin{eqnarray}
\delta\varphi^{\rm gi}_{\rm rad}(k,t)&=& a_{\rm rad}(t)^{-1}\chi_{\rm rad}(z_R),\quad\quad t\leq t_i
\label{phi-rad1}\\
\delta\varphi^{\rm gi}_{\rm inf}(k,
t)&=& a_{\rm inf}(t)^{-1}\left[C_1(k)u_{\rm inf}(z_I)+C_2(k)u^*_{\rm inf}(z_I)\right],\quad\quad t>t_i,
\label{phi-inf1}
\end{eqnarray}
where  
\begin{eqnarray}
\chi_{\rm rad}(z_R)&=&\frac{1}{\sqrt{2k}}e^{-ik\tau}=\frac{1}{\sqrt{2k}}e^{iz_R}, \\
u_{\rm inf}(z_I)&=&\sqrt{\frac{\pi z_I}{4k}}H_{\nu_\chi}^{(1)}(z_I),\\
u^*_{\rm inf}(z_I)&=&\sqrt{\frac{\pi z_I}{4k}}H_{\nu_\chi}^{(2)}(z_I),
\end{eqnarray}
and $C_1=\sqrt{\frac4\pi}\tilde c_1$ and $C_2=\sqrt{\frac4\pi}\tilde
c_2$.  A subscript $k$ for $\chi_{\rm rad}(z_R)$ and $u_{\rm
  inf}(z_I)$ is implicit. 
The Wronskian of $u_{\rm inf}(z_I)$ gives
$|C_1|^2-|C_2|^2=1$.
The task is to determine the coefficients $C_1(k)$ and $C_2(k)$.

We demand that the wavefunction of gauge invariant inflaton
fluctuation and its time derivative remain continuous at the time of the
transition, i.e.
\begin{eqnarray}
\delta\varphi^{\rm gi}_{\rm rad}(t_i)&=&\delta\varphi^{\rm gi}_{\rm inf}(t_i),\nonumber\\
\delta\dot{\varphi}^{\rm gi}_{\rm rad}(t_i)&=&\delta{\dot\varphi}^{\rm gi}_{\rm inf}(t_i).
\label{matching-condition}
\end{eqnarray}
Following Eq.~(\ref{phi-rad1}) and Eq.~(\ref{phi-inf1}) we get 
\begin{eqnarray}
\delta\dot{\varphi}^{\rm gi}_{\rm rad}(t_i)&=&\frac{\dot\chi_{\rm rad}(t_i)}{a(t_i)}-\frac{\chi_{\rm rad}(t_i)}{2t_ia(t_i)},\\
\delta{\dot\varphi}^{\rm gi}_{\rm inf}(t_i)&=&\frac{C_1(k)\dot u_{\rm inf}(t_i)+C_2(k)\dot u^*_{\rm inf}(t_i)}{a(t_i)}-\frac{C_1(k)u_{\rm inf}(t_i)+C_2(k)u^*_{\rm inf}(t_i)}{2t_ia(t_i)}.
\end{eqnarray}
Using the matching conditions given in Eq.~(\ref{matching-condition})
one gets
\begin{eqnarray}
\left.\chi_{\rm rad}\right|_{t=t_i}&=&C_1(k)u_{\rm inf}|_{t=t_i}+C_2(k)u^*_{\rm inf}|_{t=t_i},\label{chi-ti}\\
\left.\dot\chi_{\rm rad}\right|_{t=t_i}&=&C_1(k)\dot u_{\rm inf}|_{t=t_i}+C_2(k)\dot u^*_{\rm inf}|_{t=t_i}\label{chi-dot-ti},
\end{eqnarray}
where we have used Eq.~(\ref{chi-ti}) to simplify expressions and get
Eq.~(\ref{chi-dot-ti}). Alternatively we could have matched
$a\delta\varphi^{\rm gi}$ and its derivative at the transition.
Solving these two equations simultaneously and using the Wronskian
$u_{\rm inf}\dot u^*_{\rm inf}-u^*_{\rm inf}\dot u_{\rm
  inf}=\frac{i}{a(t)}$ (yielding $|C_1|^2-|C_2|^2=1$) one gets
\begin{eqnarray}
C_1(k)&=& ia(t_i)\left.\left(u^*_{\rm inf}\dot\chi_{\rm rad}-\dot u^*_{\rm inf}\chi_{\rm rad}\right)\right|_{t=t_i},\\
C_2(k)&=& ia(t_i)\left.\left(\dot u_{\rm inf}\chi_{\rm rad}-u_{\rm inf}\dot\chi_{\rm rad}\right)\right|_{t=t_i}.
\end{eqnarray}
Now to determine $C_1(k)$ and $C_2(k)$ we need $\dot\chi_{\rm rad}$,
$\dot u_{\rm inf}$ and $\dot u^*_{\rm inf}$ at $t_i$. To obtain
expressions for these three quantities we notice that $\dot z=-\frac
ka$ and the derivatives of the Hankel functions are (Eq.~(5.3.5) and Eq.~(5.4.9) of 
Ref. \cite{lebedev}):
\begin{eqnarray}
\frac{d}{dz}H^{(1,2)}_\nu(z)=\frac{\nu H^{(1,2)}_\nu(z)}{z}-H^{(1,2)}_{\nu+1}(z),
\label{derivative-hankel}
\end{eqnarray}
where $\nu$ is an arbitrary order. Using the above equations one gets
\begin{eqnarray}
\dot\chi_{\rm rad}(z_R)&=&-\frac{i}{a}\sqrt{\frac k2}e^{iz_R},\\
\dot u_{\rm inf}(z_I)&=&\frac{1}{a}\sqrt{\frac{\pi k}{4z_I}}\left[z_IH^{(1)}_{\nu_\chi+1}(z_I)-\left(\nu_\chi+\frac12\right)H^{(1)}_{\nu_\chi}(z_I)\right],\\
\dot u^*_{\rm inf}(z_I)&=&\frac{1}{a}\sqrt{\frac{\pi k}{4z_I}}\left[z_IH^{(2)}_{\nu_\chi+1}(z_I)-\left(\nu_\chi+\frac12\right)H^{(2)}_{\nu_\chi}(z_I)\right].
\end{eqnarray}
Thus the co-efficients are
\begin{eqnarray}
C_1(k)&=&ie^{iz_R}\sqrt{\frac{\pi}{8z_I}}\left.\left[\left(\nu_\chi+\frac12-iz_I\right)H_{\nu_\chi}^{(2)}(z_I)-z_IH_{\nu_\chi+1}^{(2)}(z_I)\right]\right|_{t_i},\\
C_2(k)&=&-ie^{iz_R}\sqrt{\frac{\pi}{8z_I}}\left.\left[\left(\nu_\chi+\frac12-iz_I\right)H_{\nu_\chi}^{(1)}(z_I)-z_IH_{\nu_\chi+1}^{(1)}(z_I)\right]\right|_{t_i}.
\end{eqnarray}
To check that $|C_1|^2-|C_2|^2=1$ we obtain
\begin{eqnarray}
|C_1(k)|^2-|C_2(k)|^2&=&-\frac{i\pi z_I}{4}\left.\left[H_{\nu_\chi}^{(1)}(z_I)H_{\nu_\chi+1}^{(2)}(z_I)-H_{\nu_\chi}^{(2)}(z_I)H_{\nu_\chi+1}^{(1)}(z_I)\right]\right|_{t_i}
\end{eqnarray}
Now using the relation in Eq.~(9.1.17) of Ref. \cite{abramowitz}),
\begin{eqnarray}
H^{(1)}_{\nu+1}(z)H^{(2)}_{\nu}(z)-H^{(2)}_{\nu+1}(z)H^{(1)}_{\nu}(z)=-\frac{4i}{\pi z}
\end{eqnarray}
in the above equation we get $|C_1|^2-|C_2|^2=1$.

As was discussed in \textsection \ref{sec:other_assumptions}, during inflation,
the curvature perturbation $\cal R$ receives most of its contribution from $\delta \phi$ 
and so we can use the above 
to find the primordial power spectrum. 
{
When inflation is 
preceded by a radiation era  the power spectrum for the inflaton field fluctuations in the vacuum
state is (using
arguments similar to those in Ref. \cite{Powell:2006yg})
\begin{eqnarray}
\mathcal{P}_{\delta\phi}(k)=\mathcal{P}_{\delta\phi}^{\rm BD}|C_1-C_2|^2\,,
\end{eqnarray}
where $\mathcal{P}_{\delta\phi}^{\rm BD}$ corresponds to the standard power 
spectrum in vacuum presuming Bunch-Davies initial conditions on
$\delta\varphi^{\rm gi}_{\rm inf}$.
Then
\begin{eqnarray}
\mathcal{P}_{\mathcal R}(k)=\mathcal{P}_{\mathcal R}^{\rm BD}|C_1-C_2|^2=A\left(\frac {k}{k_P}\right)^{n_s-1}|C_1-C_2|^2,
\label{powspec1}
\end{eqnarray}
where $A$ and $n_s$ are the amplitude and scalar spectral index
respectively of the spectrum in a generic inflationary scenario, and
$k_P$ is the pivot scale. Thus
the modification due to mode function-matching is a
multiplicative factor of $|C_1-C_2|^2$ which we will calculate now. We
have
\begin{eqnarray}
C_1(k)-C_2(k)&=&ie^{iz_R}\sqrt{\frac{\pi}{2z_I}}\left.\left[\left(\nu_\chi+\frac12-iz_I\right)J_{\nu_\chi}(z_I)-z_IJ_{\nu_\chi+1}(z_I)\right]\right|_{t_i},\\
C_1^*(k)-C_2^*(k)&=&-ie^{-iz_R}\sqrt{\frac{\pi}{2z_I}}\left.\left[\left(\nu_\chi+\frac12+iz_I\right)J_{\nu_\chi}(z_I)-z_IJ_{\nu_\chi+1}(z_I)\right]\right|_{t_i},
\end{eqnarray}
which yields
\begin{eqnarray}
|C_1-C_2|^2&=&\frac{\pi}{2z_I}\left[z_I^2\left(J_{\nu_\chi}^2(z_I)+J_{\nu_\chi+1}^2(z_I)\right)-2z_I\left(\nu_\chi+\frac12\right)J_{\nu_\chi}(z_I)J_{\nu_\chi+1}(z_I)\right.\nonumber\\
&&\left.\left.+\left(\nu_\chi+\frac12\right)^2J_{\nu_\chi}^2(z_I) \right]\right|_{t_i}.
\label{factor}
\end{eqnarray}
We need to evaluate $z_I(k,t_i)$ to determine the above
factor. We know that 
\begin{eqnarray}
z_I(k,t_i)=\frac{k}{(1-\epsilon_I)a_iH_i}.
\label{zi}
\end{eqnarray}
For the mode corresponding to the horizon size
at the onset of inflation we
have $k_i=a_iH_i$. 
For the largest mode of cosmological interest with wavenumber $k_{0}=a_0 H_0=H_0$ (for $a_0=1$), 
which
leaves $N(k_0)$ e-foldings before inflation ends, we have
\begin{eqnarray}
N(k_{0})-N(k_{i})=\ln(k_i/k_{0})\,.
\end{eqnarray}
This gives
\begin{eqnarray}
k_i=k_{0}e^{-\delta N},
\end{eqnarray}
where we have defined $\delta N \equiv
N(k_{i})-N(k_{0})$. Then
\begin{eqnarray}
z_I(k,t_i)=\frac{k}{(1-\epsilon_I)H_0}e^{\delta N}.
\end{eqnarray}

  
Now if we also consider a thermal distribution of the scalar field quanta
as in Sec. \ref{subseq:Thermalinitcond}
then the power spectrum will have an
additional multiplicative factor  and will be given by 
\begin{eqnarray} \label{eq:final_pps1}
\mathcal{P}_{\mathcal R}(k)=A_s^\prime\left(\frac {k}{k_P}\right)^{n_s-1}|C_1-C_2|^2\coth\left(\frac{k}{2T}\right),
\label{pi_pps1}
\end{eqnarray}
where $T$ is the comoving temperature of the scalar field quanta.
We can normalize the above equation 
as
\begin{eqnarray} \label{eq:final_pps}
\mathcal{P}_{\mathcal R}(k)=A_s\left(\frac {k}{k_P}\right)^{n_s-1}
|C_1-C_2|^2\coth\left(\frac{k}{2T}\right)
\left[|C_1(k_P)-C_2(k_P)|^2\coth\left(\frac{k_P}{2T}\right)\right]^{-1}
,
\label{pi_pps}
\end{eqnarray}
so that $P_{\cal R}(k_P) = A_s$.  
In Fig. (\ref{pps_various}) we plot the form of the Bunch-Davies
power spectrum and the modified power spectra of Eqs. (\ref{powspec1}) 
and (\ref{pi_pps}).


\begin{figure}[!htb]
\begin{center}
\includegraphics[width=6.0in]{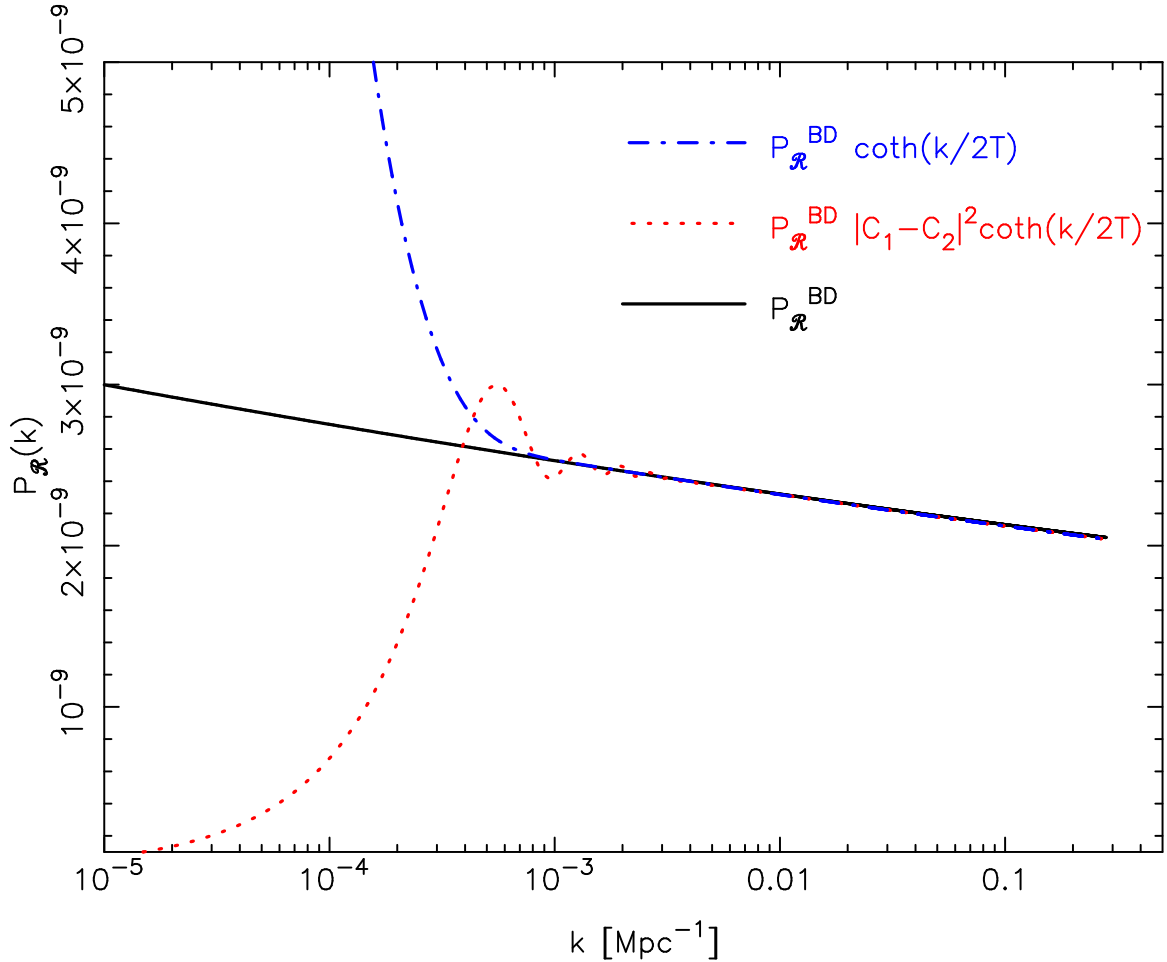}
\end{center}
\caption{The effects of various factors in Eq. (\ref{eq:final_pps}) on the shape of the scalar Primordial Power 
Spectrum. While the coth term enhances the power at low $k$ values, the factor of $|C_1-C_2|^2$ due to non-trivial 
mode dynamics lowers it. We have used $\delta N =0.081$ and 
$T=1.286\times 10^{-4}$ Mpc$^{-1}$ (from Table~(\ref{tab:wmap9planck})). 
The Bunch-Davies 
power law Primordial Power Spectrum has also been shown for reference.
}
\label{pps_various}
\end{figure}
 \subsection{Infrared divergences}

Before proceeding to the estimation of parameters for this model with a pre-inflationary radiation era, 
let us see whether the modified power spectrum is free of infrared divergences  or not.
The two-point correlation function $G$ at coincident points is
proportional to the power spectrum as 
\begin{eqnarray}
G\propto\int\frac{dk}{k}\mathcal{P}_{\mathcal R}.
\end{eqnarray}
For a generic inflationary scenario with no pre-inflationary era one has 
\begin{eqnarray}
\mathcal{P}_{\mathcal R}=A_s(k/k_P)^{n_s-1},
\end{eqnarray}
with
$n_s-1=2\eta_V-6\epsilon_V$. This spectrum diverges when $k\rightarrow0$. 
But in a scenario where the inflation is
preceded by a radiation era, the mode-matching at the boundary
brings up the multiplicative factor given in Eq.~(\ref{factor}). In the
small $k$ limit this factor will go as $\propto k^{2\nu_\chi-1}$. Now
we have $\nu_\chi\approx3/2$. Thus, in such a case the correlation
function will be
\begin{eqnarray}
G\propto\int dk\, k^{1+2\eta_V-6\epsilon_V}
\end{eqnarray}
which converges at $k\rightarrow0$ (Note that the dependence of $k$ in
such a case matches with \cite{Marozzi:2011da}). Also, if we consider
the thermal distribution of the inflaton then it will bring up an
extra factor of $\coth (2kT)$ which as $k\rightarrow0$ will go as $k^{-1}$. Thus in such a case the correlation function
will be
\begin{eqnarray}
G\propto\int dk\, k^{2\eta_V-6\epsilon_V},
\end{eqnarray}
which is also IR convergent. Hence the power spectrum is infrared divergence free once we take into account 
a radiation era preceding inflation, a fact well known in the literature (for the standard 
power spectrum without the coth factor) \cite{Ford:1977in}.

\section{Parameter estimation}  \label{sec:param_est}

In the inflationary scenario of our interest, the Primordial Power Spectrum (PPS) has two
extra parameters $ \delta N$, the number of e-foldings of inflation in excess of the standard 
minimum (which is, for example, 60 e-foldings for GUT scale inflation) and the comoving inflaton 
temperature $T$. We  try to constrain the extra parameters of PPS from the  WMAP nine year and Planck
 data in this section using Markov Chain Monte Carlo (MCMC) analysis and for this purpose 
use the publicly available code  {\tt COSMOMC} \cite{2002PhRvD..66j3511L,cosmomc}.

{\tt COSMOMC} uses a  publicly available code {\tt CAMB} \cite{2000ApJ...538..473L} based   
a line of sight integration  approach given in Ref. \cite{1996ApJ...469..437S} for 
computing the power spectra of CMB anisotropies for a set of cosmological parameters. 
Apart from  the two extra parameter $\delta N$ and $T$, characterizing the primordial 
power spectrum, we  vary  six standard parameters, namely, the physical densities of baryons 
($\Omega_bh^2$) and dark matter ($\Omega_ch^2$), the dark energy density ($\Omega_{\Lambda}$) or $\theta$ as defined later, the
amplitude  ($A_S$) and spectral index ($n_s$) of the primordial power spectrum 
(at  pivot scale $k_{P}=0.05~\text{Mpc}^{-1}$) and  the optical depth of reionization ($\tau$) in 
our MCMC analysis. 

We modify {\tt CAMB} such that it can incorporate the two extra parameters for PPS which we have.
In order to compute the likelihood  from the power spectra of the CMB anisotropies for 
WMAP nine year and  Planck data  we use the likelihood codes  provided by WMAP \cite{wmap} 
and Planck team \cite{plc} respectively. For completeness a short description of the 
WMAP nine year and Planck likelihood codes and data is given below.  

The WMAP nine year likelihood code and the methodology of parameter estimation is discussed  in detail
in Ref. \cite{2013ApJS..208...19H} and is not very different than what was outlined in Ref. \cite{2003ApJS..148..195V}.
WMAP likelihood and nine year temperature and polarization data can be 
downloaded from 
Ref. \cite{wmap}.
The code computes likelihoods for the TT, EE, TE and BB angular power spectra differently
at low and high-l.  The low-l $( l \leq 32)$  TT likelihood is computed from the power spectrum estimated
by Gibbs sampling and high-l $( l > 32)$  TT likelihood is computed using an optimal quadratic estimator. The
WMAP likelihood code computes the low-l ($l < 23$) polarization likelihood directly in
the pixel space.

At present Planck has made only the temperature data publicly available which can 
be used
alone or with a combination of other CMB data sets to constrain theoretical 
models.  The
likelihood code for Planck data  can be downloaded from Ref. \cite{plc} and a 
detailed description of
it can be found  in Ref. \cite{2014A&A...571A..15P}.
Since Planck has more frequency channels spread over a wider 
range, higher angular resolution and
better sensitivity as compared to WMAP it is far better equipped to deal
with systematics like foregrounds. Higher angular resolution and better sensitivity 
of Planck
allows us to use the angular power spectrum (temperature) up to $l=2500$, and a higher number of
frequency channels makes it  possible to model foregrounds  more accurately.

The Planck likelihood software ({\tt Clik}) has a few different likelihoods modules, some
of which are {\tt CAMspec} for computing the $TT$ likelihood at high-$l$ (up to  
$l=2500$),
{\tt commander} for computing the 
TT likelihood at low-$l$ (from $l=0$ to $l=49$) and
{\tt lowlike} for computing the low-$l$ ($l=0$ to $l=32$) polarization likelihood
from the $TT,EE,BB$ and $TE$ power spectra (for polarization it uses WMAP nine year data).

The {\tt CAMspec} module of the Planck likelihood code has 14 extra (nuisance) parameters
which are used for modeling systematics like foregrounds, asymmetric beams, etc.
In principle, these extra parameters also can be estimated from the same data set
from which the cosmological parameters are estimated.  However, we do not do that
and instead fix the values of the {\tt CAMspec} nuisance parameters to their values
reported in Refs. \cite{2014A&A...571A..15P,2014A&A...571A..16P}, and vary only the
standard six standard parameters and the extra parameters of PPS.

For our analysis we take the sum of the physical masses of the light neutrinos ($\sum_{\nu}$) as 0.6 eV, 
the effective number of neutrinos ($N_{\rm eff}$) as 3.046, the Helium mass fraction ($Y_{He}$) as 0.24 
and the width of reionization as 0.5. For the case of spatially flat background  cosmological models, 
as we consider here, either the Hubble parameter at present ($H_0$) or the dark energy density is 
considered as a fitting parameter and the other is computed from the flatness condition. In practice 
{\tt COSMOMC} does not use $H_0$ or $\Omega_{\Lambda}$ as one of the six base parameters  and rather uses $\theta$,
the ratio of the size of the sound horizon at decoupling ($r_{dec}$) and the angular diameter
distance at decoupling ($D_A$), as one of the parameters.

\begin{table}
\small 
\begin{center}
\begin{tabular}{cccl}  \hline \hline
S. No & Parameter & Prior & Description \\
1     & $\Omega_b h^2$  & $[0.005,0.1]$ &  Baryon density today \\
2     & $\Omega_c h^2$  & $[0.001,0.99]$ & Cold dark matter density today \\
3     & $100 \theta$  & $[0.55,10.0]$ & $\approx 100\times  r_{dec}/D_A$  (CosmoMC) \\
4     & $\ln [10^{10}A_s]$ & $[2.7,4.0]$ & Primordial curvature perturbations at $k_0 = 0.05\, \rm{Mpc}^{-1}$ \\
5     & $n_s$ & $[0.9,1.1]$ & Scalar spectrum power-law index at $k_0 = 0.05 \,\rm{Mpc}^{-1}$ \\
6     & $\tau$& $[0.01,0.8]$ & Thomson scattering optical depth due to reionization \\
7     & $\delta N$& $[-2.000,2.0]$  &  Number of extra e-foldings  \\
8     & $T$ & $[0.0000001,0.0004]$ & Temperature  in $\rm{Mpc}^{-1}$\\ \hline 
\end{tabular}
\caption{Cosmological parameters and the prior ranges which we use in {\tt COSMOMC}}
\label{tab:prior}
\end{center}
\end{table}

The prior ranges  which we use for parameter estimation are given in Table~(\ref{tab:prior}). 
We  found the prior ranges for the two extra parameters $\delta N$ and $T$ 
(in $\rm {Mpc}^{-1}$) by considering a few test cases.  For the rest of the cosmological parameters 
we use the same prior ranges as used in the literature \cite{2014A&A...571A..16P}.

As is clear from Eq.~(\ref{pi_pps}), the primordial power spectrum depends on
$\epsilon_I$ and $\nu_{\chi}$.
We consider two values of $\epsilon_I$, namely, $0.001$ and 0.
The choice of $\epsilon_I=0.001$ with the presumed range of $A_s$ in Table~(\ref{tab:prior}) 
effectively sets $H_i/M_{Pl}\sim10^{-6}$, or $V_i^{1/4}\sim 10^{15}$ GeV, while 
$\epsilon_I=0$ corresponds to a very low energy scale of inflation, such as at the 
electroweak scale.  
These choices are consistent with the upper bound in \textsection \ref{sec:LTE} on 
$H_i$ from the requirement of thermalization before inflation
so as to avail of the fluid approximation.
$\nu_\chi$ as defined in \textsection\ref{primpertinfl} depends on $\epsilon$ and $\eta$ during inflation.  
Since the power law factor in Eq.~(\ref{pi_pps}) is determined by the slow-roll dynamics of the inflaton,
we still have $n_s=1-4\epsilon+2\eta$ during inflation.
For the above values of $\epsilon_I$ we vary $n_s$ and can obtain 
the corresponding values of $\eta$.

For the case of $\epsilon_I = 0.001$, 
the extreme values of the prior range of $n_s$ 
according to Table~(\ref{tab:prior}), 
i.e. $0.9$ and $1.1$, lead to $\eta_V$ lying between $-0.047$ and $0.053$ which 
includes values
which are not too small as compared to the upper bound of $1/12=0.083$
in Eq. (\ref{eq:eta_V_condition}).
But the best fit value and $1\sigma$ range 
of $n_s$ (in Table (\ref{tab:wmap9planck})) 
correspond to values of $\eta_V$ that satisfy the upper bound.
We have also checked that restricting the prior range of $n_s$ to 
$[0.954,1.034]$, which corresponds to $|\eta_V| < 0.02$ (for $\epsilon_I = 0.001$)
does not change the results noticeably.
Similar remarks apply to the case when $\epsilon_I = 0$.

In order to run {\tt COSMOMC} for a theoretical model we not only need prior ranges we also need a 
covariance matrix which is used in Markov Chain Monte Carlo  sampling.  
Covariance matrices are also provided with {\tt COSMOMC}
for a large
number of theoretical models.  However,
for a new model we must find the covariance matrix by running {\tt COSMOMC} for a few test cases
and we have also done that.  

\begin{figure*}[!htb]
\begin{center}
\includegraphics[width=6.5in]{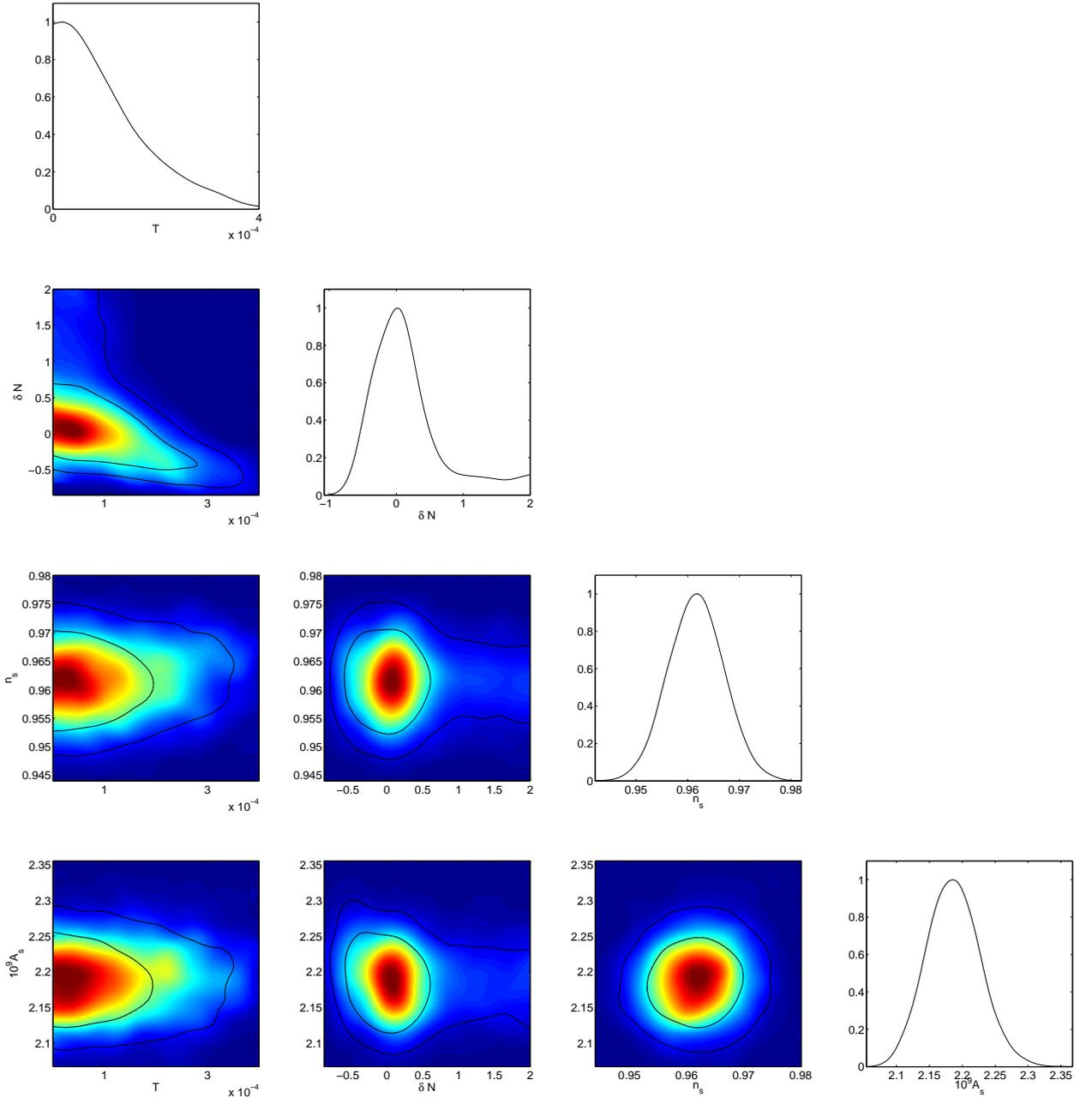}
\end{center}
\caption{The diagonal panels in this figure show the one dimensional marginalized probability 
distributions and the other panels show  the 68\% and 95\% confidence regions of the parameters for the 
WMAP nine year+ Planck  data for the pre-inflationary model for $\epsilon_I=0.001$. 
Since the rest of the cosmological parameters show the 
expected behavior  we do not show plots 
for those. 
}
\label{fig:contours_pi2_w9p}
\end{figure*}


In the diagonal panels of Fig.~(\ref{fig:contours_pi2_w9p}) we show (for the case $\epsilon_I = 0.001$)
the marginal posterior probability distributions of 
the parameters $T$, $\delta N$, $n_s$ and $A_s$ which characterize 
the primordial power spectrum, and in the other panels we show the 
68\% and 95\% confidence 
regions. We do not show plots for rest of the cosmological parameters (which have expected behavior) 
since we are primarily interested in the parameters of the primordial power spectrum. From the figure 
we can see that $\delta N$ can be well constrained using the combined WMAP nine year and Planck data, 
while we can put only an upper limit on $T$.

In Table~(\ref{tab:wmap9planck}) we present a summary of the cosmological parameters for
the standard power law model and pre-inflationary model for the joint WMAP nine year and 
Planck data for the two cases $\epsilon_I = 0.001$ and $\epsilon_I = 0$. 
Apart from showing the best fit values of the cosmological parameters for 
the power law and the pre-inflationary model,
we also give the 68\% limit for the parameters. 
As we noted above, though we can constrain $\delta N$, we can only give an upper bound on $T$. 
There is not much change in the values of the standard 
cosmological parameters and their errors when we replace the standard power law model
with the pre-inflationary model, while there is a slight 
improvement in the $-\log$ likelihood by 1.945.  However this improvement
cannot be considered 
significant, in particular when the error bars on the extra parameters $\delta N$ and $T$ 
are large. 
In Fig.~(\ref{power_w9p}), we plot the 
$TT$ angular 
power spectrum of CMB corresponding to the pre-inflationary model with
$\epsilon_I=0.001$,  and the 
power spectrum for the standard 
power law case, 
using the best fit values of parameters (and the upper bound on
$T$) from Table~(\ref{tab:wmap9planck}),
with the WMAP nine year and Planck datasets.


\begin{table*}[h]
\begin{footnotesize}
\centering
\begin{tabular}{|c|cc|cc|cc|cc|}  \hline
\multicolumn{1}{|c|}{} & \multicolumn{2}{c|}{Power Law}     &\multicolumn{2}{c|}{Pre-Inflation ($\epsilon_I=0.001$)}  &\multicolumn{2}{c|}{Pre-Inflation ($\epsilon_I=0$)}        \\  \hline
Parameter              &  Best Fit   & 68\% Limit          & Best Fit & 68\%Limit    & Best Fit    &  68 \% Limit        \\ \hline
$\Omega_bh^2$          &  0.02238    &  0.02235    $\pm$  0.00020  & 0.02234    &  0.02235    $\pm$  0.00019 &  0.02226  & 0.02235 $\pm$  0.00019   \\ \hline
$\Omega_ch^2$          &  0.1159     &  0.1165     $\pm$  0.0020  & 0.1167      &   0.1165     $\pm$ 0.0021  &  0.1171   & 0.1165  $\pm$  0.0021     \\ \hline
$\Omega_{\Lambda}$      &  0.708      &  0.704      $\pm$  0.012    & 0.702       &    0.704  $\pm$ 0.012   &  0.700  &  0.704    $\pm$  0.012    \\ \hline
$10^9A_s$              &  2.186      &  2.186      $\pm$  0.038  & 2.178        &   2.186     $\pm$   0.040 &  2.163  &  2.187    $\pm$  0.039       \\ \hline
$n_s$                  &  0.962      &  0.962      $\pm$  0.005 &  0.961      &  0.961    $\pm$   0.005    &  0.960  & 0.962     $\pm$  0.005   \\ \hline
$\tau$                 &  0.089      &  0.089      $\pm$  0.009 &  0.087      & 0.089      $\pm$  0.009    &  0.083  & 0.089     $\pm$  0.010       \\ \hline
$H_0$                  &  69.04     &  68.74       $\pm$  0.97    & 68.58       & 68.73      $\pm$ 0.96  &  68.40  & 68.75     $\pm$  0.98    \\  \hline
$\delta N$             &             &                              & 0.081  &  0.160       $\pm$   0.549   & 0.017   &    0.171 $\pm$   0.552       \\ \hline
$T$                    &             &                           &      &   $ < 1.2867 \times 10^{-4}  $   &    &    $<  1.2782 \times 10^{-4}$         \\ \hline
-$\log {\mathcal L}$   & 8691.8200   &                       &          & 8689.8750    &                     & 8689.9980          \\ \hline
\end{tabular}
\caption{Best fit cosmological parameters estimated from the WMAP nine year + Planck data 
for the power law model, and the pre-inflationary model with $\epsilon=0.001$
and 0 with $\delta N$ and $T$ as extra parameters.  
(For $T$ we only obtain upper bounds.)
Because of skewed non-gaussian distributions the values in the 3rd, 5th 
and 7th columns
are not centred about the best fit values.
}
\label{tab:wmap9planck}
\end{footnotesize}
\end{table*}

\begin{figure}[!htb]
\begin{center}
\includegraphics[width=6.0in]{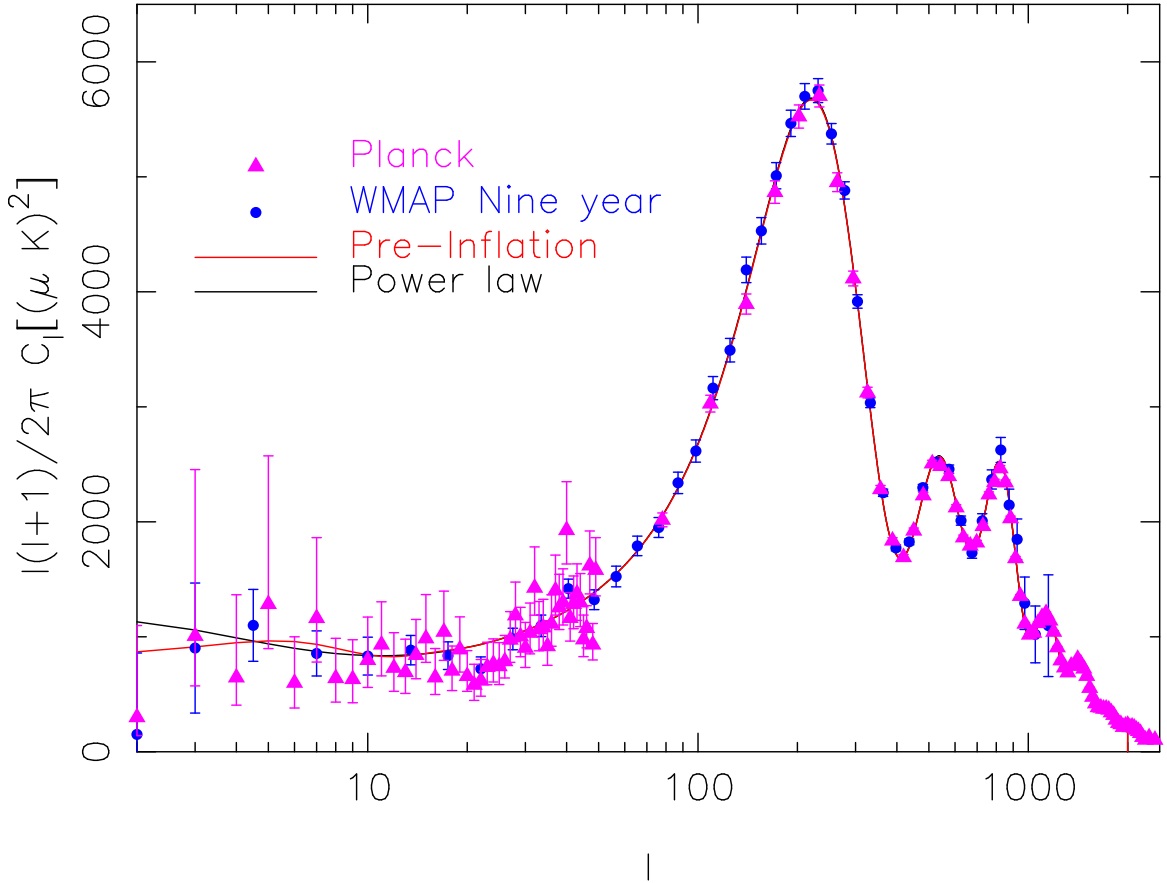}
\end{center}
\caption{We show the  $C_l^{TT}$ for Planck (pink filled 
triangles) and the  WMAP nine year data
(large blue dots). The black and red curves respectively show the 
best-fit theoretical $C_l^{TT}$ for the standard power law model 
and the pre-inflationary model with $\epsilon_I=0.001$
using values from Table~(\ref{tab:wmap9planck}).  For the latter curve,
$\delta N=0.081$ and $T=1.2867\times 10^{-4}$ Mpc$^{-1}$. 
}
\label{power_w9p}
\end{figure}

\section{Discussion and Conclusions} \label{sec:conclusions}

Considering a radiation dominated era prior to inflation we have
found the primordial power spectrum of adiabatic scalar 
perturbations 
as given by Eq.~(\ref{pi_pps}) and shown in Fig.~\ref{pps_various}. 
In the previous literature where a pre-inflationary radiation era has been
discussed,
either a modification of the mode functions (due to the presence of a prior radiation era) is considered
\cite{Powell:2006yg,Marozzi:2011da,Wang:2007ws,Hirai:2002vm,Hirai:2003dh,
Hirai:2004kh,Hirai:2005tn,Hirai:2007ne}, 
or, a thermal distribution of inflaton quanta (which carries a signature of their prior thermal equilibrium) has been considered \cite{2006PhRvL..96l1302B}. 
While the first tends to lower the quadrupole moment of the CMB $TT$ anisotropy spectrum, the 
latter enhances the power at large angular scales. We have 
included in this study both these phenomena 
as a consistent approach to determine the physics of the subsequent inflationary era. 
In Fig.~\ref{pps_various} where we plot the Primordial Power Spectrum with 
the $|C_1-C_2|^2$ and the coth term 
one sees that the damping of the power on large scales due to $|C_1-C_2|^2$ overwhelms the enhancement 
due to the $\rm {coth}(k/(2T)]$ factor.
(The ringing behaviour in the power spectrum seen in Fig.~\ref{pps_various} is associated with the abrupt transition from the radiation to the inflationary era at $t_i$ -- though $a$ and $H$ are continuous at $t_i$, $\epsilon$ is not.  This behaviour is also seen in Fig. 1 of Ref. \cite{Powell:2006yg} and has been discussed in Ref. \cite{Wang:2007ws}.  In Refs. \cite{2014PhRvD..89b4024G,2015PhRvD..91b4014G} particle creation associated with the abrupt transition in $\epsilon$ between eras has been regulated by letting the Bogolyubov parameter $\beta_k\rightarrow \beta_k e^{-\bar{\tau}k}$ for some very small
but finite time duration $\bar \tau$ of the transition.  If one tries to consider $\epsilon$ varying over a short time scale during the transition between eras then it will not be possible to obtain analytical solutions for the equations for the mode functions.)

Using WMAP nine year and Planck data, we find that 
the upper bound on the comoving temperature of inflaton is given by 
$T \lesssim 1.28 \times 10^{-4} {\rm Mpc}^{-1}$, and the best fit value of the minimum 
number of e-folds is $0.081$ more than the minimum value required to solve the horizon and flatness problems
(see Table \ref{tab:wmap9planck} and Fig. \ref{fig:contours_pi2_w9p}). 
These results do not change significantly for the smaller energy scale of inflation 
$(\epsilon_I=0)$. 
We have thus improved the existing constraint on the comoving temperature of the inflaton particles 
from $T \lesssim  10^{-3} {\rm Mpc}^{-1}$ (according to Ref. \cite{2006PhRvL..96l1302B})
to  $T \lesssim  10^{-4} {\rm Mpc}^{-1}$, i.e. by one order of magnitude.
\footnote{
If we repeat the analysis of Ref. \cite{2006PhRvL..96l1302B}, i.e. without modifications of the mode functions, and include only one extra parameter $T$ in COSMOMC then, with the newer WMAP9 + Planck data, we again find $T\lsim 10^{-4}{\rm Mpc}^{-1}$.  This will imply an increase of 2 in the minimum $\delta N$ of $7-32$ for GUT-electroweak scale inflation obtained in Ref. \cite{2006PhRvL..96l1302B}. ($\delta N$ is large in this analysis because there is no suppression of the coth factor by the $|C_1-C_2|^2$ factor.)
}

Our analysis shows that considering the effect of both the modified mode functions
and the temperature of the inflaton quanta on the primordial power spectrum
marginally improves the likelihood compared to the standard case though, as mentioned
earlier, its significance is diminished because of the large error bars for $\delta N$ and $T$. 
We also depict the CMB power spectrum in the presence of a pre-inflationary radiation era 
with the 
WMAP nine year and Planck data in 
Fig.~\ref{power_w9p}.
It is evident from the plot that the best fit spectrum tends to lower the quadrupole 
moment. 
However the decrease is not significant enough for us
to claim that the low $l$ anomaly present in the Planck data can 
indicate a pre-inflationary radiation era. 

One motivation of having a pre-inflationary radiation era is to get rid of the 
infrared divergences which
appear in the field theoretic treatment of perturbations in the 
inflationary scenario. We have shown that the corrected power spectrum obtained 
in our analysis is still infrared safe, even with
the coth term, and hence curing infrared divergences with
the existence a pre-inflationary era
holds good in our scenario as well. 
We have also discussed the conditions under which the radiation dominated era
prior to inflation can be treated in the fluid approximation, as needed in the
standard cosmological perturbation theory.

\begin{acknowledgments}
Numerical work for the present study was done on the IUCAA HPC facility. Work of SD is supported by Department of Science and
 Technology, Government of India under the Grant Agreement number IFA13-PH-77 (INSPIRE Faculty
 Award). JP would like to thank the Science and Engineering Research Board (SERB) of the Government of India for  financial support  via a Start-Up Research Grant (Young Scientists)  SR/FTP/PS-102/2012. SD would like to thank Tarak Thakore for help in numerical analysis. GG thanks Subhendra Mohanty for discussions.  GG and RR would like to thank Namit Mahajan for useful suggestions.
 SD and RR would like to thank Anjishnu Sarkar for discussions in the early part of the collaboration and also acknowledge the organizers of WHEPP XI, which was
 held at the Physical Research
 Laboratory, Ahmedabad, for providing a stimulating environment for ideas which ultimately led to this article.

\end{acknowledgments}

\bibliographystyle{JHEP}
\bibliography{pi}

\end{document}